# On the economics of knowledge creation and sharing

Omar Metwally, MD
omar.metwally@gmail.com
University of California, San Francisco

First Draft: September 11th 2017

## Abstract

This work bridges the technical concepts underlying distributed computing and blockchain technologies with their profound socioeconomic and sociopolitical implications, particularly on academic research and the healthcare industry. Several examples from academia, industry, and healthcare are explored throughout this paper. The limiting factor in contemporary life sciences research is often funding: for example, to purchase expensive laboratory equipment and materials, to hire skilled researchers and technicians, and to acquire and disseminate data through established academic channels. In the case of the U.S. healthcare system, hospitals generate massive amounts of data, only a small minority of which is utilized to inform current and future medical practice. Similarly, corporations too expend large amounts of money to collect, secure and transmit data from one centralized source to another. In all three scenarios, data moves under the traditional paradigm of centralization, in which data is hosted and curated by individuals and organizations and of benefit to only a small subset of people.

# 1. Introduction

In its current siloed state, data is a liability rather than an asset. The value of data depends on its quantity and quality. Organizations, including corporations, government, and academia, have few incentives to share data outside the context of selling it. For instance, advertisers use data procured from individuals' browsing history and social media use (via internet service providers, social media and search engines) to create detailed profiles of individuals' online behavior and spending habits and more effective sell products to unknowing consumers. While this paradigm fits naturally into a capitalistic society, these economics of data collection and transfer do not facilitate the generation or sharing of knowledge in the academic setting.

A typical university-based research group depends upon external funding to support its research activities. These funds often originate from governmental bodies, philanthropic organizations, or corporations and are difficult to secure [1]. Only a small minority of tenure track scientists ever becomes principal investigators, and a lab that is productive today can become defunct tomorrow if its principal investigator is unable to secure funding for laboratory equipment and supplies such as microscope parts, reagents, and to compensate technicians and trainees [2]. Principal investigators spend a majority of their time writing grant applications rather than participating directly in the process of knowledge generation [3].

It is often said that publications are the currency of academia. The maxim "publish or perish" applies to most research groups, whose work culminates in peer-reviewed publications with publication fees commonly amounting to several thousand dollars [4]. Moreover, these peer-reviewed publications are heavily biased toward so-called "positive results," in which mathematical correlations between variables are described [5].

The vast majority of data produced by scientific researchers do not refute the null hypothesis; in a best case scenario, they are deemed "negative results," and are discarded; in a worst case scenario, they are data that can't be replicated, verified, or are outright fraudulent [6]. The result is the modern-day academic machinery. This severely flawed system, a victim of many conflicting economic forces, results in a tremendously inefficient workflow in which most grant money is wasted in the form of negative, and therefore unpublishable, results. Principal investigators spend a majority of their time trying to secure funding. The ultimate winner is the $10 billion business of academic publishing [6]. In this reality, data with the potential to produce vast knowledge is rendered into a vastly wasted opportunity to exponentially build on communities' resources. Individuals' roles are minimized by the centralization of resources in the hands of a privileged few.

## 2. Background

While the term "blockchain" has been touted to near-hysteria in popular media in the context of initial coin offerings and get-rich-quick schemes, an understanding of this data structure's logic reveals the tremendous and fascinating socioeconomic implications of storing data on blockchain. In its most simplified form, a blockchain is a ledger [7]. The reason for blockchain's natural association with financial derivatives lies in its ability to mathematically prove the authenticity of data and demonstrate proof of stake and proof of work [8].

The starting port for these use cases is the typical consumer, who is separate from (and often completely unaware of) the data collected about him or her. For instance, a customer's online behavior is collected and used to up-sell the customer as much as algorithmically possible [9]. Customers have nothing to gain (and a few thousand dollars each year in extra spending to lose) from such data, which companies can sell to data brokers and merchants [10]. Analogously, the majority of taxpayers have no access to — and oftentimes no way to directly benefit from — publications funded through research that ends up property of academic journals [11, 12].

**2.1 Case Study: Proof of Stake**

Consider a research lab living from grant to grant, sifting through negative results to find crumbs of publishable positive results. If its lab notebooks were stored in the form of a blockchain, every experiment conducted, every machine learning model and dataset, and every clinical trial would generate data that lives on the blockchain as a cryptographic asset. Also referred to as "coins" and "tokens," these cryptographic assets have inherent value because they are perfect receipts of the existence and transfer of data [13]. Never before in history has such a perfect ledger existed [14, 15]. On the blockchain, a relatively worthless set of negative results generated by a research lab becomes, when combined with negative results from thousands of other research groups, a trove of extremely valuable scientific data which can be traced to its owners whenever and however it is utilized. This large collection of negative results can become the source of unexpected positive results.

Moreover these blockchain-hosted data take on a new life as a financial derivative [16, 17]. These cryptographic assets, perfect receipts of the creation and movement of knowledge, can be traded by third-parties analogously to the way a company's common stock is bought and sold on private and public marketplaces, albeit without the same regulations and on a different scale [13]. These tokens enable individuals, small and large groups alike to be compensated for their services in ways that are impractical or impossible in traditional economies [18]. Rather than relying on the slow and ineffi-cient process of securing funding through grants, research labs can codify contracts on the blockchain to allow third-parties to bid for services and products rendered, on the metadata (what kind of knowledge research labs generate through their scholarly activities), and allow third-parties to become stakeholders in a research group's success by directly benefitting from these research activities. For instance, if I believe that a particular group is contributing to science and society in a positive way, I can economically support this group by donating computing power and electrical energy to support the integrity of their lab notebook-turned-ledger, or by trading fiat for tokens representing proof of stake in their scholarly activities. What are today opportunities exclusive to accredited investors and institutions will become abundant opportunities for individuals to influence how perceived value circulates through society.

**2.2 Case Study: Proof of Work**

Consider the United States healthcare system, which still excludes millions of Americans from access to healthcare and financially ruins even more [19, 20]. Insurance companies are able to impose high premiums simply because they can. This is the logic of a capitalistic society, and insurance companies alone enjoy the benefits of owning valuable health data to their fullest extent — at the expense of those whose health data was collected [20, 21]. Imagine, on a smaller scale, a radiology group that puts a copy of every imaging study they do on a blockchain, along with a timestamp, a description of which type of study was done, and why it was performed. In doing so, data that would have otherwise been discarded can be engaged with by third-parties while directly benefiting the radiology group as well. For instance, grassroots-based health insurance co-ops could emerge from these sources of data which are otherwise privy to insurance companies, to the benefit of health consumers, who can undergo imaging studies and receive other other healthcare services at a fraction of current costs. Information about which studies are performed — where, by whom, and why, and the result of those studies, can be used to lower healthcare costs while improving health outcomes, rather than raise healthcare costs and increasing profits.

One question that naturally arises, especially in the context of current centralized data paradigms, is: why would healthcare providers be incentivized to make public valuable data that is routinely used by corporations and insurance companies to maximize profits? One powerful force driving healthcare costs upward is the process through which health providers bill patients via insurance companies. Whether ordering relatively common drugs or expensive therapeutics or procedures, healthcare systems rely on administrators whose role is to submit authorization requests to insurance companies for approval to prescribe therapeutics on their patients' behalf [22]. When a service is rendered in the hospital or in a clinic, a healthcare team is reimbursed a fraction of the amount they bill for, creating a cat and mouse game in which providers continuously bill as high as possible for services rendered with the expectation that they will only receive a fraction of what they bill for, and in which insurance companies place limitations on which drugs and services this will pay for and how much of the cost they will cover [23]. Blockchain would provide an end to this cat-and-mouse game and create a race to the bottom for healthcare costs, through price transparency and elimination of bloated administrative layers that handle authorization requests and billing, while creating a race to the top for healthcare outcomes as this ledger of health services and outcomes would be publicly accessible on a blockchain. Simultaneously, healthcare providers can immediately receive payment for services rendered, and although individual payments may be less, overall profits would increase because payments would arrive immediately and there would be no need for entire departments of administra-tors whose entire role is to maximally inflate bills sent to insurance companies (and patients, insured and uninsured) and to see these bills through collection.

**2.3 Informing current and future medical practice**

We may well already have all the knowledge we need to cure many illnesses currently considered incurable [24]. We may well have all the data we need to create intelligent machines that can interpret CT scans, diagnose disease, and synthesize drugs to cure any condition. The reason this knowledge hasn't culminated in more rapid advancement in healthcare and science is that information is fragmented into pieces, siloed, and ultimately rendered worthless data. Blockchain allows transparent access to data. It would be naive to imply that a data structure will cure society of all its ailments. However blockchain allows data to culminate into extremely valuable information, once at the disposal of a powerful few, now to the benefit of all who become stakeholders by contributing to, interacting with, and propagating data.

# 3. The need for a ledger of scholarly assets

The need for this project, a protocol for the hosting and sharing of data on a distributed network ("Distributed Data Sharing Hyperledger," or *DDASH)*, arises from the observations by the above examples, as well as the observation that numerous research groups at UCSF and other academic institutions are working in parallel in their endeavors to create knowledge with little synergistic interaction [25]. How would research group A at UCSF Medical Center know that research group B at the University of Michigan is working to answer the same scientific questions, for instance? Without a transparent glimpse into which resources an organization owns and how they are being used and

Page 7 of 19

shared, both research groups miss opportunities for synergistic collaboration, within and among organizations.

Those acquainted with the politics of contemporary academia will be quick to raise several criticisms. Working within the current reality of Google, the most comprehensive collection of information known to humanity as of September 2017, why can't research groups A and B simply host their digital assets — data and knowledge gleamed from this data — on websites or public databases? And if groups A and B are competing to be the first to publish in academic journals and competing to drink from the same pools of grant funding, why would any research group benefit by sharing the results of experiments that were costly to run before they can reap the benefits of publication and intellectual property [26]? The answer is in blockchain's ability to capture proof of work and proof of stake in a network's digital assets. There is nothing to stop a competing research group from stealing these data and benefiting at their competitors' expense. Hosting data in the form of knowledge on a blockchain elegantly solves this problem through irrefutable mathematical proof of data ownership, transfer, and veracity [27].

**3.1 Distributed Data Sharing Hyperledger (DDASH)**

DDASH is a ledger of scholarly data and knowledge produced by life science, informatics, and clinical researchers at UCSF and other academic institutions. The need for this project arises from the negative impact of data siloing, competition, and counterproductive financial incentives in the academic world on the creation and sharing of knowledge. Concretely, researchers can host data — datasets, experimental results, and machine learning models, among other examples of scholarly knowledge — on the distributed InterPlanetary Filesystem (IPFS) network and record the location of these assets on an Ethereum-based blockchain, along with a description of the asset, when it was created, and who has privileges to access the data.

### 3.2 Network Architecture

We believe that the IPFS protocol's combination of security and speed is well suited for this application. IPFS uses content-based addressing, in which a hashing function determines a file's network address based on the file's contents [28]. Storing data in the form of a directed acyclic diagram (in this case, a Merkle DAG) results in trees that can be efficiently traversed and queried. IPFS is a peer-to-peer network in which data is continuously circulating through network participants' machines which are running the client software. Data are rendered permanent by virtual of content-based addressing and persistent by virtue of its peer-to-peer architecture, and data are rapidly accessible without the bottlenecks that Internet Protocol imposes.

### 3.3 Blockchain as a ledger

The blockchain functions as a decentralized ledger of digital resources and the movement of these resources throughout the network. As the DDASH protocol is formalized, more robust mechanisms for associating IPFS hashes with the owner of the resource and the permissions granted by the owner are necessary. Currently the DDASH protocol accounts for the following elements:

- IPFS content-addressed hash, which defines the location of an asset on the IPFS network
- The owner's public key fingerprint

- The public key fingerprints of users authorized to access the resource, or a designation as "public"
- Timestamp

In its current form, DDASH interfaces between the IPFS network and the Ethereum blockchain. One can conceive an alternative version of the DDASH protocol that seamlessly integrates a ledger-based indexing and permission management system, using for example IPFS's native public and private keys and a native IFS ledger. Keeping the networking architecture separate from the blockchain has tangible advantages, however, including the versatility of allowing users to create digital assets using any permutation of blockchains, private and public.

**3.4 Security**

DDASH allows users to manage access to privileged resources using public-key encryption. Public-key encryption allows users to identify themselves on the network using a verifiable public key, which can be used to encrypt resources such that they can only be unencrypted using a corresponding private key accessible exclusively to the intended recipient. Future versions of the DDASH protocol may feature ways to host resources on private clusters and manage access to these clusters on the blockchain. In doing so, resources are secured by limiting the movement of certain data to a subset of the swarm (network peers), and through a second layer of encryption. This not only allows data to move much more quickly through a network, it also greatly enhances security compared to the antiquated paradigm of data hosted on centralized, and therefore inherently vulnerable, servers. Common sources of wasted IT budgets and wasted productivity, such as forgotten, cracked and stolen passwords, or easily-intercepted HTTP network traffic, are obviated by virtue of the DDASH protocol. What stands between the theoret-ical underpinnings of this protocol and its implementation in academic centers and healthcare systems is not a question of the feasibility of this technology, but rather, whether legislation governing health information and computing will keep up with emerging trends in computing. Catastrophic beaches of sensitive consumer information, such as the Equifax data breach, have become regular occurrences and urgent reminders of the shortcomings of our antiquated Internet Protocol and undeserved trust in institutions that centralize large amounts of highly sensitive data at individuals' expense [29].

**3.5. DDASH Repository**

DDASH is hosted as an open source repository at https://github.com/osmode/ddash.

We intend for this nascent project to illustrate the concepts and the larger vision outlined here while serving as a starting point for a formalized protocol for hosting and interacting with distributed digital assets. We made this a public repository early in the conception of this project in order to allow the codebase to benefit from the technical expertise and creativity of the open source community, and to allow the project to benefit from the rapid and exciting evolution in computing paradigms driven by the blockchain and distributing computing communities.

## 4. Using DDASH

DDASH currently runs on the *blackswan* private Ethereum network at 104.236.141.200. It benefits from the open source work produced by
the IPFS, Ethereum, OpenPGP, web3.py, and py-ipfs communities.

The Go Ethereum client, *web3.py,* and *py-ipfs* Python packages are all prerequisite. The instructions here are for machines running Ubuntu 16.04. A Ethereum node must be connected to the blackswan private network and possess the ability to lock/unlock accounts to send transactions.

### 4.1 Directory Structure

Start by creating these directories:

```
mkdir /home/omarmetwally/blackswan
mkdir /home/omarmetwally/blackswan/gnupg
mkdir /home/omarmetwally/blackswan/data
```

### 4.2 Genesis Block

To connect to the blackswan network, you'll need to use the same genesis block defined in genesis.json (see the Github repository). Move this file to `/home/omarmetwally/blackswan/` and set your genesis block (you only need to do this once, and you need to install the Ethereum go client geth and Ethereum developer tools first):

```
geth --datadir=/home/omarmetwally/blackswan/data init /home/omarmetwally/blackswan/genesis.json
bootnode --genkey=boot.key
bootnode --nodekey=boot.key
```

### 4.3 Go Ethereum client and IPFS daemons

In order to use the *web3.py* and *ipfs* wrappers, you'll need to run *geth* and ipfs daemons in the background, respectively:

```
geth --verbosity 1 --datadir /home/omarmetwally/blackswan/data --networkid 4828 --port 30303 --rpcapi="db,eth,net,web3,personal,web3" --rpc 104.236.141.200 --rpcport 8545  console
```

Be very careful when enabling RPC while your accounts are unlocked. This can lead to Ethereum wallet attacks, hence the recommendation to keep your development environment completely separate from any real Ether you might own.

The above command starts the go Ethereum client on your local machine and attempts to connect to the blackswan server at 104.236.141.200. Remember to set your genesis block according to the above directions. Trying to join this network with a different genesis block (such as the default genesis block) will not work.

Then open a new terminal window or tab and start the ifps daemon:

```
ipfs daemon
```

### 4.4 DDASH command line interface

Once your Ethereum and IPFS nodes are running, your account is unlocked, and you can interact with both clients, start the DDASH command line interface (CLI):

```
python main.py
```

```
         _____   _____          _____  _    _
        |  __ \ |  __ \   /\   / ____|| |  | |
        | |  | || |  | | /  \ | (___  | |__| |
        | |  | || |  | |/ /\ \ \___ \ |  __  |
        | |__| || |__| / ____ \____) || |  | |
        |_____/ |_____/_/    \_\_____/|_|  |_|

    ::: Distributed Data Sharing Hyperledger :::
    https://github.com/osmode/ddash

    Welcome to the DDASH Command Line Interface.
```

[1]    ddash> sanity check
       IPFS and geth appear to be running.
[2]    ddash> set directory /home/omarmetwally/blackswan/gnupg
[3]    ddash> new key
[4]    ddash> show keys
[5]    ddash> use key 0
[6]    ddash> show accounts
[7]    ddash> use account 0
[8]    ddash> set recipient your_recipient's_pubkey_id
[9]    ddash> set file /path/to/clinical/trial/data.csv
[10]   ddash> encrypt
[11]   ddash> upload
[12]   ddash> checkout QmUahy9JKE6Q5LSHArePowQ91fsXNR2yKafTYtC9x-QqhwP

The above commands:

1. check if IPFS daemon and Go Ethereum client are running

2. specify working directory (need to have read/write permission)

3. generate a new PGP keypair

4. list all PGP keypairs on your machine

5. uses the first (index 0) keypair as your identity

6. list Ethereum accounts

7. specify index of Ethereum account to use for transactions

8. specify an intended recipient's public key

9. upload the file to IPFS and create transaction containing the hash, user id of the person who uploaded the file, and recipient's public key id (or "public" indicating that it's not encrypted).

10. encrypt file from step 9 using public key from step 8

11. upload file from step 9 to IPFS network

12. check blockchain using IPFS hash as handle

**4.5 Mining on the *blackswan* Ethereum network**

Mining difficulty is currently relatively easy (1e6) on the blackswan network. Mine Ether by running:

```
geth --verbosity 4 --datadir /Users/omarmetwally/Desktop/black-
swan/data --networkid 4828 --port 30303 --rpc 104.236.141.200--
rpcport 8545  --mine console
```

## 5. Acknowledgements

I'm grateful to my mentor, Dr. David Avrin (UCSF) for his belief in this vision and for his unwavering support. My colleagues, Dr. Michael Wang and Dr. Steven Chan, provided formative feedback during the conception of these ideas. Steven Truong (UC Berkeley) inspired me with his technical creativity. Visionaries such as Vitalik Buterin and Juan Benet, and many brilliant minds contributing to the open source communities they inspired, conceived the technical underpinnings which are allowing these concepts to grow into powerful tools which I believe will transform and modernized academic research.

20. Metwally O. "Building smart contract-based health insurance." Published: 30 June 2014. https://omarmetwally.wordpress.com/2014/06/30/building-smart-contract-based-health-insurance/.

21. Angrisano C et al l (McKinsey Global Institute). "Accounting for the cost of health care in the United States." Jan 2017. http://www.mckinsey.com/industries/healthcare-systems-and-services/our-insights/accounting-for-the-cost-of-health-care-in-the-united-states.

22. California Department of Health and Human Services. "Treatment Authorization Request." http://www.dhcs.ca.gov/provgovpart/Pages/TAR.aspx.

23. Jiwani A et al. "Billing and insurance-related administrative costs in the United States' health care: synthesis of micro-costing evidence." BMC Health Serv Res. 2014 Nov 13;14:556. doi: 10.1186/s12913-014-0556-7.

24. Hamermesh RG and Guisti K. "One obstacle to curing cancer: patient data isn't shared." Harvard Business Review, 28 Nov 2016. https://hbr.org/2016/11/one-obstacle-to-curing-cancer-patient-data-isnt-shared.

25. Distributed Data Sharing Hyperledger. https://github.com/osmode/ddash.

26. *Fecher B, Friesike S and Hebing M.* "What drives academic data sharing?" PLoS One Published: February 25, 2015. Available: https://doi.org/10.1371/journal.pone.0118053.

27. Ethereum Foundation. "A next-generation smart contract and decentralized application platform." https://github.com/ethereum/wiki/wiki/White-Paper .

28. Benet J. "IPFS - content addressed, versioned, p2p file system." arXiv: 1407.3561 [cs.NI].

29. Gressen S (Federal Trade Commission). "The Equifax Data Breach: What to Do." Published: 8 September 2017. Available: https://www.consumer.ftc.gov/blog/2017/09/equifax-data-breach-what-do
Page 19 of 19